\documentclass[journal,article,accept,moreauthors,pdftex,12pt,a4paper]{mdpi}
\lastpage{x}
\doinum{10.3390/------}
\pubvolume{xx}
\pubyear{2015}
\history{Received: xx / Accepted: xx / Published: xx}

\Title{Inhomogeneous Dark Fluid and Dark Matter, Leading to a Bounce Cosmology}

\Author{I. Brevik $^{1,}$* and A. V. Timoshkin $^{2}$}

\address{%
$^{1}$ Department of Energy and Process Engineering, Norwegian University of Science and Technology, N-7491 Trondheim , Norway \\
$^{2}$ Tomsk State Pedagogical University, ul. Kievskaya, 60, 634061 Tomsk, Russia}

\corres{E-mail: iver.h.brevik@ntnu.no, Tel.: +47-7359-3555, Fax.: +47-7359-3491.}

\abstract{The purpose of this short review is to describe cosmological  models with a linear inhomogeneous time-dependent equation of state (EoS) for the dark energy, when the dark fluid is coupled with dark matter. This may lead to a bounce cosmology. We consider equivalent descriptions in terms of the EoS parameters for an exponential, a power-law, or a double-exponential law for the scale factor $a$. Stability issues are discussed by considering small perturbations around the critical points for the bounce, in the early as well as in the late, universe. The latter part of the paper is concerned with dark energy coupled with dark matter in viscous fluid cosmology. We allow the bulk viscosity $\zeta=\zeta(H,t)$ to be a function of the Hubble parameter and the time, and consider the Little Rip, the Pseudo Rip, and the bounce universe.  Analytic expressions for characteristic properties of these cosmological models are obtained. }


\begin{document}


\section{Introduction}

The current procedure for explaining the observed expansion for the universe \cite{riess98,perlmutter99} is to introduce a dark energy fluid with negative pressure and negative entropy \cite{peebles03,sahni00,li11}. According to the astronomical observations the dark energy currently accounts for about 73\% of the total mass/energy of the universe and only 27\% of a combination of dark matter and baryonic matter \cite{kowalski08}. The dark energy universe may have very interesting implications in the future \cite{nojiri05}. A recent review of dark energy is given in Ref.~\cite{bamba12}. A different way of accounting for the dark energy without any extra components is the modification of gravity (for  recent reviews, see Ref.~\cite{nojiri11,capozziello11}). Note that in such a case there is the possibility of a unified description of early-time inflation and late-time acceleration as first realized in the Nojiri-Odintsov model \cite{nojiri07,nojiri03} of $F(R)$ gravity (for more models of unified inflation with dark energy in modified gravity, see Ref.~\cite{nojiri11}). In fact, the inhomogeneous fluid cosmology may also represent specific versions of $F(R)$ gravity \cite{nojiri07}.

In the coupled phantom/fluid model, dark energy and dark matter are usually described by assuming an ideal (non-viscous) fluid with an unusual equation of state (EoS). New forms for the EoS for general (inhomogeneous/imperfect) dark fluid models were considered in Refs.~\cite{nojiri05a,nojiri06,brevik04,brevik07,brevik13,capozziello06}.
The equation of state dark energy parameter  $w$ is known to be negative:
\begin{equation}
w=\frac{p_D}{\rho_D}<0, \label{1}
\end{equation}
where  $\rho_D$  is the dark energy and  $p_D$  is the dark pressure. According  to present observational data the value of $w$ is $w=-1.04^{+0.09}_{-0.10}$   \cite{nakamura10}. Various possible scenarios for the evolution of the universe discussed in the literature are concerned with the so-called Big Rip \cite{caldwell03,nojiri03a}, the Little Rip  \cite{frampton11,brevik11,frampton12,astashenok12,astashenok12a,astashenok12b,nojiri,makarenko13}, the Pseudo Rip \cite{frampton12a} and the Quasi Rip \cite{wei12} cosmologies.

On the other hand in the early universe there is the possibility of a matter bounce cosmological model (see Refs.~\cite{brandenberger10,cai12,xue10} and references therein).
The universe goes from an  accelerated collapse era to an expanding era over the bounce without displaying a singularity, what implies  a cyclic universe. After the bounce the universe soon
enters a phase of matter-dominated expansion. Recently the $F(R)$ and $F(T)$ gravity models in the presence of a bounce cosmology were discussed in Refs. \cite{bamba14,astashenok13,bamba14A,odintsov14A,bamba14B}. We will in the following study similar bounce cosmologies.

 We will study a cosmological system with two coupled fluids: a dark energy component with a linear inhomogeneous EoS, and a dark matter component with a linear homogeneous EoS in a flat homogeneous and isotropic FRW universe. There exist other investigations dealing with the coupling between the inhomogeneous fluid dark energy and dark matter components \cite{brevik13a,timoshkin10}.
We will explore  bounce cosmological models where  the scale factor is described by exponential as well as  power-law forms and realize them in  terms of the parameters of the EoS for the dark fluid. Then we will consider the stability of these models in the early universe by considering perturbations to the first order. Also we will consider a bounce cosmological model when the  scale factor has a double exponential form. The bouncing behavior, both in the early universe and in the late-time universe,  will be presented. The stability conditions for this model are discussed.

For an introduction to bounce cosmology in general, the reader may also consult Ref.~\cite{novello08}.

 In  Section 4 we  study examples of Little Rip, Pseudo Rip, and bounce cosmology, described in flat Friedmann-Robertson-Walker spacetime when the cosmic fluid is  viscous and is coupled with dark matter. We   find corrections to the thermodynamical parameter, and to the bulk viscosity, in the equation of state for the dark energy.

\section{	Bounce cosmology via an inhomogeneous fluid}

In this section we  construct  bounce cosmological models induced by  coupled dark energy in  terms of the parameters in the equation of state.

Let us consider a universe filled with two interacting ideal fluids: a dark energy component and a dark matter component in a spatially flat Friedmann-Robertson-Walker metric with  scale factor $a$. The background equations are given by \cite{nojiri05}:

\begin{equation}
\left\{ \begin{array}{lll}
\dot{\rho}+3H(p+\rho)=-Q, \\
\dot{\rho}_m+3H(p_m+\rho_m)=Q, \\
\dot{H}=-\frac{k^2}{2}(p+\rho+p_m+\rho_m),
\end{array} \label{2}
\right.
\end{equation}				
where   $H=\dot{a}/a$ is the Hubble rate and  $k^2=8\pi G$  with $G$ denoting Newton's gravitational constant;  $p,\rho$  and  $p_m,\rho_m$  are the pressure and the energy density of dark energy and  dark matter correspondingly;  $Q$ is a function that accounts for the energy exchange between  dark energy and  dark matter. Here a dot denotes  derivative with respect to cosmic time $t$.
The Friedman equation for the Hubble rate is given by \cite{nojiri05}:
\begin{equation}
H^2=\frac{k^2}{3}(\rho+\rho_m). \label{3}
\end{equation}
We will investigate the cases where the scale factor has an exponential, a power-law, or a double exponential form, and study bouncing behavior around $t=0$.

\subsection{Exponential model}

Let us first assume a  bounce cosmological model where the scale factor $a$ has   an exponential form \cite{bamba14}:
\begin{equation}
a=\exp(\alpha t^2), \label{4}
\end{equation}
$\alpha$ being a positive constant. The  instant $t=0$ ($a=1$) is taken to be the instant when bouncing occurs, in accordance with earlier treatments (cf., for instance, Ref.~\cite{bamba14}).  The Hubble parameter becomes
\begin{equation}
H=2\alpha t. \label{5}
\end{equation}
We suppose that the dark energy component obeys an inhomogeneous EoS \cite{nojiri05a}:
\begin{equation}
p=w(t)\rho+\Lambda(t). \label{6}
\end{equation}
 The dark matter component is taken to obey a homogeneous EoS \cite{bamba12}:
 \begin{equation}
 p_m=\tilde{w}\rho_m. \label{7}
 \end{equation}
We will in the following assume that the dark matter is dust matter, $p_m=0$, corresponding to $\tilde{w}=0$.

Now consider the gravitational equation of motion for the dark matter:
\begin{equation}
\dot{\rho}_m+3H{\rho}_m=Q. \label{8}
\end{equation}
We choose the interaction term between  dark energy and  dark matter in the following form, depending exponentially on time:
\begin{equation}
Q(t)=Q_0\exp \left(-\frac{3}{2}Ht\right), \label{9}
\end{equation}
with $Q_0$ a constant. Thus $Q(0)=Q_0$ when  $t=0$. The solution of Eq.~(\ref{8}) becomes
\begin{equation}
\rho_m(t)=(\rho_0+Q_0t)\exp \left( -\frac{3}{2}Ht\right), \label{10}
\end{equation}
where $\rho_m(0)=\rho_0$. According to this the dark matter energy increases linearly with $t$ for low $t$, and fades away when $t\rightarrow \infty$.

Taking into account Eqs.~(\ref{2}, \ref{3}) we obtain the gravitational equation of motion
\begin{equation}
\frac{6H\dot{H}}{k^2}-\dot{\rho}_m+3H\left[ (1+w)\left(\frac{3}{k^2}H^2-\rho_m\right)+\Lambda \right]=-Q. \label{11}
\end{equation}
As the time development of  dark matter   is known from Eq.~(\ref{10}), Eq.~(\ref{11}) can be regarded as the equation of motion for the dark energy.

Let us choose the parameter $w(t)$ in the form
\begin{equation}
w(t)=-1-\frac{\delta k^2}{3}H^2, \label{12}
\end{equation}
 with $\delta$ a positive constant. Thus $w(t)<0$ always, corresponding to the phantom region. When $t\rightarrow 0$, $w(t)\rightarrow -1$ (the cosmological constant case), while when $t\rightarrow \infty$, $w(t)\rightarrow -\infty$.

 Solving Eq.~(\ref{11}) with respect to $\Lambda(t)$ (the time-dependent cosmological 'constant') when $w(t)$ is taken to have the form (\ref{12}),  we obtain
 \begin{equation}
 \Lambda(t)= -\frac{4\alpha}{k^2}+\delta H^4-\left(w+\frac{3}{2}\right) \rho_m. \label{13}
 \end{equation}
Thus  when $t\rightarrow 0$ one sees that $\Lambda \rightarrow -4\alpha/k^2-\rho_0/2$ (cf. Eq.~(\ref{10})), while when $t\rightarrow \infty$, $\Lambda(t)\rightarrow \delta (2\alpha t)^4 \rightarrow \infty$.

 Concluding this subsection, we have  explored the exponential model of bounce cosmology in terms of time-dependent parameters in the EoS, taking into account the interaction between dark energy and dark matter.

 \subsection{Power-law model}

 We next consider the case where the scale factor has the following form \cite{bamba14} (see also \cite{timoshkin10}):
 \begin{equation}
 a=\bar{a}\left(\frac{t}{\bar{t}}\right)^q+1, \label{14}
 \end{equation}
 where $\bar{a} \neq 0$ is a constant, $\bar{t}$ is a reference time, and $q=2n, n \in Z$.

 In this case the Hubble parameter becomes
 \begin{equation}
 H=\frac{\bar{a}(q/\bar{t})( t/\bar{t})^{q-1}}{\bar{a}(t/\bar{t})^q+1}. \label{15}
 \end{equation}
 In analogy to the case above, we assume also now that the dark matter pressure is zero, $p_m=0$.

 The derivative of  $H$ with respect to cosmic time   is equal to
 \begin{equation}
 \dot{H}=H\left[ \frac{t^{-1}(q-1)}{\bar{a}(t/\bar{t})^q+1}-\frac{H}{q}\right]. \label{16}
 \end{equation}
 Let us choose the interaction term between dark energy and dark matter in the form
 \begin{equation}
 Q=\frac{Q_0}{\left[\bar{a}(t/\bar{t})^q+1\right]^3}. \label{17}
 \end{equation}
 As before, $t=0$ means the bouncing time, corresponding to $Q(t=0)=Q_0$. When $t\rightarrow \infty$, the interaction term $Q(t)\rightarrow 0$.

 Solving the gravitational equation of motion (\ref{8}) for the dark matter, we find
 \begin{equation}
 \rho_m(t)=\frac{\rho_0+Q_0t}{\left[ \bar{a}(t/\bar{t})^q+1\right]^3}, \label{18}
 \end{equation}
 where $\rho_m(0)=\rho_0$.

 Let us assume that the thermodynamic parameter  $w(t)$ in the EoS (\ref{6}) has the same form (\ref{12}) as before. Taking into account Eqs.~(\ref{15})-(\ref{18}) we obtain from Eq.~(\ref{11}) the following expression for the cosmological 'constant':
 \begin{equation}
 \Lambda(t)=\delta H^4-\frac{2H}{k^2}\left[ \frac{t^{-1}(q-1)}{\bar{a}(t/\bar{t})^q+1}-\frac{H}{q}\right] -\left[w-\bar{a}\left(\frac{t}{\bar{t}}\right)^q\right]\rho_m. \label{19}
 \end{equation}
Near $t=0$ for $n\neq 1$  we  have $\Lambda \rightarrow \rho_0$. In the case $n=1$ we have $\Lambda \rightarrow -4\bar{a}/k^2\bar{t}^2+\rho_0$.

As a result we have explored how the power-law expression (\ref{14}) for the scale factor, in addition to the form (\ref{17}) for $Q$ and the form (\ref{12}) for $w(t)$,  influence the  parameter  $\Lambda(t)$ in the EoS for the bouncing universe.

\subsection{Double exponential model}

 Now we will consider a double exponential model where the scale factor has the form \cite{bamba14}	
 \begin{equation}
 a=\exp(Y)+\exp(Y^2), \label{20}
 \end{equation}				
  where   $Y=(t/\bar{t})^2$ and  $\bar{t}$  is as before a reference time. In this model there is a unification of the bouncing behavior both in the early universe and in the late-time cosmic acceleration.

The Hubble parameter is equal to
\begin{equation}
H=\frac{2}{\bar{t}}\sqrt{Y}\,\frac{\exp(Y)+2Y\exp(Y^2)}{\exp(Y)+\exp(Y^2)}. \label{21}
\end{equation}
We suppose that the interaction term $Q$ has the form
\begin{equation}
Q=\frac{Q_0}{a^3}, \label{22}
\end{equation}
with $Q=Q_0$ at $t=0$.

The we can find the solution of the gravitational equation of motion for the dark matter \cite{nojiri05a}:
\begin{equation}
\rho_m(t)=\frac{\rho_0}{a^3}+Q\bar{t}\,\sqrt{Y}, \label{23}
\end{equation}
from which
\begin{equation}
\dot{\rho}_m(t)=Q-3H\rho_m(t). \label{24}
\end{equation}
If we take the parameter $w(t)$ to have the same form (\ref{12}) as before, we can solve $\Lambda(t)$ from Eq.~(\ref{11}):
\begin{equation}
\Lambda(t)=\delta H^4-\frac{2}{k^2}\left\{ \frac{H}{\bar{t}\sqrt{Y}}
-H^2+\left( \frac{2}{\bar{t}}\right)^2Y\left[ 1+\frac{(1+4Y^2)\exp(Y^2)}{a}\right] \right\}+w\rho_m. \label{25}
\end{equation}
When  $t\rightarrow 0$ the cosmological 'constant' $\Lambda \rightarrow -2/(k^2\bar{t})-\rho_0$.

As result we have constructed the double exponential model in the terms of the EoS parameters  for the coupled dark energy.

\section{Stability of the solutions}
			
In this section we  examine the stability of the solutions obtained in the previous section. First of all, we establish the formalism of this method. The evolution equations (\ref{2}) can be expressed as a plane autonomous system \cite{nojiri05}
\begin{equation}
\left\{ \begin{array}{lll}
\frac{dx}{dN}=-[1+p'(\rho)]\left[3+\frac{Q}{2H\rho_k}\right]x+3x[2x+(1+w_m)(1-x-y)], \\
\frac{dy}{dN}=-[1-p'(\rho)]\left[3x+\frac{Q}{2H\rho_v}y\right]+3y[2x+(1+w_m)(1-x-y)], \\
\frac{1}{H}\frac{dH}{dN}=-\left[3x+\frac{3}{2}(1+w_m)(1-x-y)\right].
\end{array} \label{26}
\right.
\end{equation}
here we have defined nondimensional quantities,
\begin{equation}
x=\frac{k^2\rho_k}{3H^2}, \quad y=\frac{k^2\rho_v}{3H^2}, \label{27}
\end{equation}
where $\rho_k$ and $\rho_v$ are the "kinetic" and "potential" terms,
\begin{equation}
\rho_k=\frac{1}{2}(\rho+p), \quad \rho_v=\frac{1}{2}(\rho-p), \label{28}
\end{equation}
and $N=\ln a$, $p'(\rho)=dp/d\rho$.

We shall investigate the stability of the critical points in the bounce models. Let us consider the early universe ($t\rightarrow 0$), where bounce cosmology can be realized. We assume for simplicity that the bounce takes place at $t=t_*$, so that $t*<0$ $(t_*>0)$ represents the contracting (expanding) phase. The bounce time $t_*$ is assumed to be very small. In this model the universe has a collapsing era for $t<t_*$, and thereafter initiates an expanding era.

With these assumptions the autonomous system (\ref{26}) for the exponential, the power-law, and the double exponential models can be written on the common form
\begin{equation}
\left\{ \begin{array}{ll}
\frac{dx}{dN}=3x(1+x-y), \\
\frac{dy}{dN}=3[y(1+x-y)-2x].
\end{array} \label{29}
\right.
\end{equation}
Setting  $dx/dN=0$ and $dy/dN=0$ in Eqs.~(\ref{29}), one obtains for the  critical points: $(x_c', y_c')= (0,0)$  and  $(x_c'', y_c'')=(0,1).$

Consider now small perturbations $u$ and $v$ around the critical point $(x_c.y_c)$, i.e.,
\begin{equation}
x=x_c+u, \quad y=y_c+v. \label{30}
\end{equation}
With $(x_c',y_c')=(0,0)$ the substitution of (\ref{30}) into (\ref{29}) leads in the linear approximation to the first order differential equations
\begin{equation}
\left\{ \begin{array}{ll}
\frac{du}{dN}=3u, \\
\frac{dv}{dN}=3(v-2u).
\end{array} \label{31}
\right.
\end{equation}
The general solution of (\ref{31}) for the evolution of linear perturbations can be written as
\begin{equation}
\left\{ \begin{array}{ll}
u=c_1\exp(3N), \\
v=(c_2+c_3N)\exp(3N),
\end{array} \label{32}
\right.
\end{equation}
where $c_1,c_2$ and $c_3$ are arbitrary constants.

In the limit $N\rightarrow \infty$, $u,v \rightarrow \infty$. There occurs an asymptotic instability of the solution around this critical point. This instability may be a consequence of of the influence from dark matter on dark energy.

 In the case  $(x_c'',y_c'')=(0,1)$ the autonomous system (\ref{29}) in the linear approximation becomes
 \begin{equation}
 \left\{ \begin{array}{ll}
 \frac{du}{dN}=0, \\
 \frac{dv}{dN}=-3(u+v).
 \end{array} \label{33}
 \right.
 \end{equation}
The solution of (\ref{33}) has the form
\begin{equation}
\left\{ \begin{array}{ll}
u=c_1, \\
v=c_2+c_3\exp(-3N).
\end{array} \label{34}
\right.
\end{equation}
In the case $c_3=0$ we obtain a one-parametric family of points at rest on a straight line. The resting point $(0,1)$ is stable.

This concludes our study of stability of the solutions of the bounce cosmological models.

\section{Dark energy coupled with dark matter in viscous fluid cosmology}

Our second theme of the present review is to describe the theory of dark energy when it is coupled with dark matter in viscous cosmology. Cosmological models treating dark energy and dark matter as imperfect fluids with an unconventional EoS were considered in Refs.~\cite{nojiri05a,capozziello06,nojiri06}. In these cases a viscous fluid is just one particular option. As is known, the case of a nonviscous  (ideal) fluid is an idealized model, which is useful in many instances but in some situations incorrect.  Viscosity, for example, is an essential ingredient when considering turbulence effects in cosmology \cite{brevik12a}. From a hydrodynamic viewpoint the viscosity effect describes the deviation from thermodynamic equilibrium to the first order. The influence from bulk viscosity in the cosmic fluid plays an important role for the Big Rip singularity \cite{brevik06a,nojiri05a}, or in type II, III and IV Rip singularities \cite{nojiri05a,brevik10a}. The Little Rip and Pseudo Rip phenomena and the bounce cosmology  interpreted as an inhomogeneous nonviscous dark fluid coupled with dark matter, are considered in Refs.~\cite{brevik13a,brevik14a}. Cosmological models in which the  modification of gravity is described in terms of a viscous fluid are explored in Refs.~\cite{myrzakul14} and \cite{myrzakulov14a}.

A key reference for the material in the following is our recent paper \cite{brevik14b}.

\subsection{Little Rip }

Among the theories of future singularities the most dramatic one is  that describing the Big Bang scenario, in which the universe encounters a singularity in a finite time. There are also milder variants of the future singularity phenomenon, namely the so-called Little Rip and the Pseudo Rip. We  now review these scenarios briefly.

 First consider the Little Rip. It is characterized by an energy density $\rho$ increasing with time,  but in such a way that an {\it infinite time} is required to reach the  singularity. It corresponds to an equation-of-state parameter $w<-1$, but $w \rightarrow -1$ asymptotically.

\bigskip

Let us take the   Hubble parameter to increase exponentially with time, cf.  \cite{frampton12},
\begin{equation}
H=H_0e^{\lambda t}, \quad H_0>0, \lambda>0, \label{3A}
\end{equation}
where $H_0=H(0)$, $t=0$ denoting present time.

Assume the  dark matter to be dust,  $p_m=0$, so that  the gravitational equation for dark matter reduces to
\begin{equation}
\dot{\rho}_m+3H\rho_m=Q. \label{4A}
\end{equation}
With the same coupling as in \cite{nojiri11},
\begin{equation}
Q=\delta H\rho_m, \label{5A}
\end{equation}
the solution of Eq.~(\ref{4A}) for dark matter becomes
\begin{equation}
\rho_m(t)=\tilde{\rho}_0\exp \left(\frac{\delta-3}{\lambda}H \right), \label{6A}
\end{equation}
where $\tilde{\rho}_0$ is an integration constant. If $\delta$ is small ($<3$),
$\rho_m \rightarrow 0 $ when $t\rightarrow \infty$.

One can choose different forms for the equation of state for the viscous fluid. We adopt
 the  inhomogeneous form; cf.  \cite{myrzakul14}
\begin{equation}
p=w(\rho)\rho-3H\zeta(H), \label{7A}
\end{equation}
implying  that the bulk viscosity depends on the Hubble parameter. Further, we choose $w(\rho)$
 to have the form; cf. \cite{myrzakul14}
\begin{equation}
w(\rho)=A_0\rho^{\alpha-1}-1, \label{8A}
\end{equation}
where $A_0 \neq 0$ and $\alpha \geq 1$ are constants.  This leads to the  the gravitational equation of motion for  dark energy
\begin{equation}
\frac{6H^3}{k^2}-\dot{\rho}_m+3H[A_0\rho^\alpha -3H\zeta(H)]=-\delta H\rho_m. \label{9A}
\end{equation}
From this  we obtain an explicit expression for the Hubble-dependent bulk viscosity
\begin{equation}
\zeta(H)=\frac{2}{3}\frac{\lambda}{k^2}+\frac{1}{3H}(\rho_m+A_0\rho^\alpha ),
\label{10A}
\end{equation}
 where  the second term contains the contribution from the coupling.

 When $t \rightarrow \infty$, $H\rightarrow \infty$ exponentially, as does the density $\rho$.   The second term to the right in Eq.~(\ref{10A}) goes to infinity in this limit.

It is of interest to consider also another variant of the Little Rip model where the Hubble parameter increases with time according to a double exponential,
\begin{equation}
H=H_0\exp(Ce^{\lambda t}), \label{11A}
\end{equation}
where $H_0, C$ and $\lambda$ are positive constants.

 We take the interaction term $Q$ to have the same form (\ref{5A}) as before.
Solving the gravitational equation of motion (\ref{4A}) for dark matter, we find
\begin{equation}
\rho_m (t)=\tilde{\rho}_0\exp \left( \frac{e^C}{C}\,\frac{\delta-3}{\lambda}H\right). \label{12A}
\end{equation}
Here $\tilde{\rho}_0$ is an integration constant.

We keep the same form (\ref{8A}) for the dark energy thermodynamical parameter $w(\rho)$ in Eq.~(\ref{7A}) as before. Then we obtain from the gravitational equation of motion (\ref{9A}) for dark energy
\begin{equation}
\zeta(H,t)=\frac{2C\lambda}{3k^2}e^{\lambda t}+\frac{A_0\rho^\alpha}{3H}+\frac{\rho_m}{3H}\left[ \frac{\delta}{3}(1-e^C)+e^{C+\lambda t} \right]. \label{13A}
\end{equation}
In this case the cosmic time occurs explicitly in the viscosity, not only indirectly via $\rho, \rho_m$ and $H$.

\subsection{Pseudo Rip }

This is another  soft variant of the future singularity scenario in which the Hubble parameter
 tends to a cosmological {\it constant} as $t\rightarrow \infty$. That means, the universe approaches a de Sitter space. We will now analyze this model in analogy to the analysis of the Little Rip model.

\bigskip

Assume first that the Hubble parameter has the form; cf. \cite{frampton12}
\begin{equation}
H=H_0-H_1\exp(-\tilde{\lambda}t), \label{14A}
\end{equation}
where $H_0, H_1$ and $\tilde{\lambda}$ are positive constants, $H_0>H_1, t>0$.
Thus in the  late-time universe $H\rightarrow H_0$.

With the interaction between dark energy and dark matter in the form (\ref{5A}), we find as solution of the gravitational equation (\ref{4A})
\begin{equation}
\rho_m(t)=\tilde{\rho}_0\exp \left[ (\delta-3)\left(H_0t-\frac{H-H_0}{\tilde{\lambda}}\right) \right], \label{15A}
\end{equation}
with $\tilde{\rho}_0$  an integration constant.

If the thermodynamical parameter $w(\rho)$ has the form (\ref{8}) we obtain from Eq.~(\ref{9A})  the bulk viscosity as
\begin{equation}
\zeta(H,t)=\frac{2\tilde{\lambda}}{3k^2}\left( \frac{H_0}{H}-1\right)+\frac{1}{3H}(\rho_m+A_0\rho^\alpha ), \label{16A}
\end{equation}
showing again the coupling.

\bigskip

Let us also consider, as a second example,   a cosmological model where the
 Hubble parameter has the form; cf. \cite{frampton12}
\begin{equation}
H=\frac{x_f}{\sqrt 3 }\left[ 1-\left(1-\frac{x_0}{x_f}\right) \exp \left( -\frac{\sqrt{3}At}{2x_f}\right)\right]. \label{17A}
\end{equation}
 Here $x_0=\sqrt{\rho_0}$ corresponds to the present energy density, $x_f$ is a finite, and $A$ is a positive constant. In the late universe   $H\rightarrow x_f/\sqrt 3$, and the expression (\ref{17A}) tends asymptotically to the de Sitter solution. When $t\rightarrow 0$, we obtain $H\rightarrow  x_0/\sqrt 3$.

For the energy density of dark matter we find
\begin{equation}
\rho_m(t)=\tilde{\rho}_0\exp \left\{ (\delta-3)\frac{x_f}{\sqrt 3}\left[ t+\frac{2x_f}{\sqrt{3}A}\left[1-\frac{\sqrt 3}{x_f}H\right)\right]\right\}, \label{18A}
\end{equation}
with $\tilde{\rho}_0$  an integration constant.

In this case we derive the following   bulk viscosity:
\begin{equation}
\zeta(H,t)=\frac{A}{3k^2H}\left( 1-\frac{\sqrt 3}{x_f}H\right)+\frac{1}{3H}(\rho_m+A_0\rho^\alpha). \label{19A}
\end{equation}
The interaction between dark energy and dark matter implies corrections in the bulk viscosity. This contrasts the case where only dark energy was involved.

\section{	Conclusion	}
						
In the present paper we  investigated  bounce cosmological models in which we  took into account the  interaction between  dark energy  and dark matter.   In terms of  parameters of the equation of state, $w(t)$ and $\Lambda(t)$ (cf. Eq.~(\ref{6})), we  described  bounce cosmologies when the scale factor is expressed by either an exponential, a power-law, or a  double exponential form. In all cases, we  adopted the form (\ref{12}) for the parameter $w(t)$. It turned out that near the bouncing instant $t=0$, the expressions for $\Lambda(t)$ reduced essentially to the initial density of dark matter, $\rho_0$.

Not very much is known from observations about the form of the interaction term $Q$. The analytic forms given above, in Eqs.~(\ref{9}), (\ref{17}), and (\ref{22}), were motivated chiefly by mathematical tractability. Physically, we have as checking points that $Q\rightarrow 0$ when $t\rightarrow \infty$.

 We also  analyzed in a linear  approximation the stability of the stationary points against perturbations and  showed the existence of a stable point, and attractor solutions, for these models.

In the second part of our paper we  studied examples of Little Rip, Pseudo Rip, and bounce cosmology, described in flat Friedmann-Robertson-Walker spacetime when the cosmic fluid is  viscous and is coupled with dark matter. We found corrections in the thermodynamical parameter, and to the bulk viscosity, in the equation of state for the dark energy.

 It could be  mentioned that there are common properties of the corrections present in these models. Thus $\Delta \zeta_m(H,t)=\rho_m/(3H)$ represents the correction to the bulk viscosity from the coupling. Further, $\Delta \zeta_\rho(H,t)=(A_0/3H)\rho^\alpha$ is related to the choice of thermodynamical parameter $w(\rho)$ in the equation of state.

From a general perspective  the present theory   may be considered as inhomogeneous cosmic fluid theory with changeable thermodynamical parameter $w$ coupled to dark matter \cite{elizalde14}, applied to the Little Rip, the Pseudo Rip, and the bounce phenomena.

As is known, cf. \cite{nojiri11},  a viscous fluid may be understood also as a modified gravity model,
for instance of the $F(R)$ type. It is moreover known that $F(R)$ gravity may provide a unification of early-time inflation with a special version of dark energy, as was proposed in Ref.~\cite{nojiri03}, or with the nonlinear model proposed in Ref.~\cite{nojiri08}.     Having that in mind, we expect that there is a natural possibility to unify these epochs with inflation in an extended viscous model.

There are different types of  bounce universes. We have presented some examples that are in conformity with our model. Other types of bounce universes, such as those encountered in Matter Bounce Loop Quantum Cosmology for instance, will be discussed elsewhere.

\bigskip

{\bf Acknowledgments}

\bigskip

We thank Professor V. V. Obukhov for valuable collaboration in a number of related works.

This  work  is supported by the Ministry of Education and Science of Russian Federation(VVO,AVT).


 The authors declare no conflict of interest.

\bibliographystyle{mdpi}
\makeatletter
\renewcommand\@biblabel[1]{#1. }
\makeatother



%


%

\end{document}